\documentstyle[12pt]{article}

\catcode`\@=11
\long\def\@makefntext#1{
\protect\noindent \hbox to 3.2pt {\hskip-.9pt  
$^{{\ninerm\@thefnmark}}$\hfil}#1\hfill}                

\def\@makefnmark{\hbox to 0pt{$^{\@thefnmark}$\hss}}  
        
\def\ps@myheadings{\let\@mkboth\@gobbletwo
\def\@oddhead{\hbox{}
\rightmark\hfil\ninerm\thepage}   
\def\@oddfoot{}\def\@evenhead{\ninerm\thepage\hfil
\leftmark\hbox{}}\def\@evenfoot{}
\def\sectionmark{}\def\subsectionmark{}}

\renewcommand{\thefootnote}{\fnsymbol{footnote}}

\renewcommand{\subsubsection}[1]
{\vspace*{0.6cm}\addtocounter{subsubsectionc}{1}
        \noindent
{\normalsize\rm\thesectionc.\thesubsectionc.\thesubsubsectionc. 
        #1}\par\vspace*{0.6cm}}

\newcounter{appendixc}
\newcounter{subappendixc}[appendixc]
\newcounter{subsubappendixc}[subappendixc]

\renewcommand{\appendix}[1] {\vspace*{0.6cm}
        \refstepcounter{appendixc}
        \setcounter{figure}{0}
        \setcounter{table}{0}
        \setcounter{equation}{0}
        \renewcommand{\thefigure}{\Alph{appendixc}.\arabic{figure}}
        \renewcommand{\thetable}{\Alph{appendixc}.\arabic{table}}
        \renewcommand{\theappendixc}{\Alph{appendixc}}
        \renewcommand{\theequation}{\Alph{appendixc}.\arabic{equation}}
        \noindent{\bf Appendix \theappendixc #1}\par\vspace*{0.4cm}}

\def\abstracts#1{{
       
\centering{\begin{minipage}{12.2truecm}\footnotesize\baselineskip=12pt\noindent
        \centerline{\footnotesize ABSTRACT}\vspace*{0.3cm}
        \parindent=0pt #1
        \end{minipage}}\par}} 

\renewenvironment{thebibliography}[1]
        {\begin{list}{\arabic{enumi}.}
        {\usecounter{enumi}\setlength{\parsep}{0pt}
\setlength{\leftmargin 1.25cm}{\rightmargin 0pt}
         \setlength{\itemsep}{0pt} \settowidth
        {\labelwidth}{#1.}\sloppy}}{\end{list}}

\newcounter{itemlistc}
\newcounter{romanlistc}
\newcounter{alphlistc}
\newcounter{arabiclistc}

\newcommand{\fcaption}[1]{
        \refstepcounter{figure}
        \setbox\@tempboxa = \hbox{\footnotesize Fig.~\thefigure. #1}
        \ifdim \wd\@tempboxa > 6in
           {\begin{center}
        \parbox{6in}{\footnotesize\baselineskip=12pt Fig.~\thefigure. #1}
            \end{center}}
        \else
             {\begin{center}
             {\footnotesize Fig.~\thefigure. #1}
              \end{center}}
        \fi}

\newcommand{\tcaption}[1]{
        \refstepcounter{table}
        \setbox\@tempboxa = \hbox{\footnotesize Table~\thetable. #1}
        \ifdim \wd\@tempboxa > 6in
           {\begin{center}
        \parbox{6in}{\footnotesize\baselineskip=12pt Table~\thetable. #1}
            \end{center}}
        \else
             {\begin{center}
             {\footnotesize Table~\thetable. #1}
              \end{center}}
        \fi}

\def\@citex[#1]#2{\if@filesw\immediate\write\@auxout
        {\string\citation{#2}}\fi
\def\@citea{}\@cite{\@for\@citeb:=#2\do
        {\@citea\def\@citea{,}\@ifundefined
        {b@\@citeb}{{\bf ?}\@warning
        {Citation `\@citeb' on page \thepage \space undefined}}
        {\csname b@\@citeb\endcsname}}}{#1}}

\newif\if@cghi
\def\cite{\@cghitrue\@ifnextchar [{\@tempswatrue
        \@citex}{\@tempswafalse\@citex[]}}
\def\citelow{\@cghifalse\@ifnextchar [{\@tempswatrue
        \@citex}{\@tempswafalse\@citex[]}}
\def\@cite#1#2{{$\null^{#1}$\if@tempswa\typeout
        {IJCGA warning: optional citation argument 
        ignored: `#2'} \fi}}

 1
 1
 1

\font\ninerm=cmr9

\def\lsim{\mathrel{\mathop{\kern 0pt <}\limits_{\displaystyle\sim}}}
\def\gsim{\mathrel{\mathop{\kern 0pt >}\limits_{\displaystyle\sim}}}
\begin{document}

\centerline{\normalsize\bf FIXED POINTS AND POWER CORRECTIONS\footnote
{Talk presented at the
'Int. Workshop on Deep Inelastic Scattering and Related Phenomena' (DIS96),
Rome, Italy, April 15-19, 1996.
}}
\baselineskip=22pt

\vspace{0.6cm}
\centerline{\footnotesize Georges GRUNBERG}

\baselineskip=13pt
\centerline{\footnotesize\it Centre de Physique Th\'eorique de l'Ecole
 Polytechnique\footnote{CNRS UPRA 0014}}
\baselineskip=12pt
\centerline{\footnotesize\it 91128 Palaiseau Cedex - France}
\vspace{0.6cm}
\centerline{\footnotesize E-mail: grunberg@orphee.polytechnique.fr}

\vspace{0.9cm}
\abstracts{\normalsize The connection between renormalons and power corrections
is discussed in the case the effective coupling constant has an infrared fixed
point of perturbative origin.}
\vspace{10cm}
CPTh/PC463.0896
\hspace{9cm}
August 1996

\vspace{0.5cm}
\newpage 
\pagestyle{plain}
\normalsize\baselineskip=15pt
\setcounter{footnote}{0}
\renewcommand{\thefootnote}{\alph{footnote}}
The connection between renormalons and Landau singularity has been pointed out
 recently~\cite{{g},{du}} to be more subtle then usually believed . In
particular ,  
it has been shown that renormalons are still present even in the case the 
effective coupling constant has an infrared (IR) fixed point of entirely 
perturbative 
origin ,  and hence no Landau singularity for positive initial values of the 
coupling.That the position and nature  of the 
renormalon is determined only by the first two terms (one and two 
loop) of the beta function was known long ago~\cite{{p},{m}}(in this sense
, the 
renormalon is a perturbative singularity , although its normalization depends on
all orders of perturbation theory~\cite{{g2},{bz}}).This result implies however 
that the Borel 
sum,which is ambiguous owing to the renormalon,  cannot coincide with the exact
amplitude ,which is well defined in this case:they in fact differ by a (complex)
power correction.I shall review a simple toy model where this fact is
explicitly 
demonstrated.Consider the typical IR renormalon integral:

\begin{equation} R(\alpha) = \int_{0}^{Q^{2}} n \frac{dk^2}{k^2}
\left(\frac{k^{2}}
{Q^{2}}
\right)^{n}
\alpha_{eff} (k/Q,\alpha)\end{equation}
where $\alpha = \alpha_{eff}(k=Q)$ , and $\alpha_{eff}(k)$ is a renormalization 
group (RG) invariant effective 
coupling
(I assume $n > 0$, so that the integral in Eq.~(1) is IR convergent order by 
order 
in perturbation theory).Assume further that $\alpha_{eff}(k)$ satisfies the two
loop renormalization group equation:
\begin{equation} \frac{d\alpha_{eff}}{d ln k^2} = - \beta_0 (\alpha_{eff})^2 - 
\beta_1 (
\alpha_{eff})^3\end{equation}
Performing the change of variable (adapted from a similar one suggested 
in~\cite{ma}): 

\begin{equation}\frac{z}{z_n} = \frac{1 - \frac{\alpha}{\alpha_{eff}(k)}}{1 + 
\frac{\beta_1}
{\beta_0} \alpha}\end{equation}
(with $z_n = n/\beta_0$) , 
and assuming that  $\beta_1/\beta_0 < 0$ , so that $\alpha_{eff}(k)$ has an 
IR fixed point at $\alpha_{IR} = -\beta_0/\beta_1$,the integral in Eq.~(1) 
becomes~\cite{g}: 
\begin{equation} R(\alpha) = \int_0^{z_n} dz\ exp\left(-
\frac{z}{\alpha}\right) 
\frac{exp\left(-\frac{\beta_1}{\beta_0}\ z\right)}{\left(1 - \frac{z}{z_n}
\right)^{1+\delta}}\end{equation}
where  $\delta = \frac{\beta_1}{\beta_0} z_n$ . Eq.~(4) differs from the Borel 
sum : 
\begin{equation} R_{PT}(\alpha) \equiv 
\int_0^{\infty} dz\ exp\left(- \frac{z}{\alpha}\right) 
\frac{exp\left(-\frac{\beta_1}{\beta_0}\ z\right)}{\left(1 - \frac{z}{z_n}
\right)^{1+\delta}} \end{equation}
by a power correction:
\begin{eqnarray} R_{POW}(\alpha) = - \int_{z_n}^{\infty} dz\ exp\left(-\frac{z}
{A} \right) 
\frac{1}{\left(1 - \frac{z}{z_n}\right)^{1+\delta}} 
& = & - \tilde{C}\ exp\ (-z_n/A)(-1/A)^{\delta}\nonumber\\
& = & - \tilde{C}(-1)^{\delta}\ (\Lambda^2/Q^2)^n\end{eqnarray}
where $1/A = 1/\alpha + \beta_1/\beta_0$ ,  $\tilde{C} = 
\frac{\beta_0}{\beta_1} \Gamma(1-\delta)(z_n)^{\delta}$ and the solution of
 Eq.~(2) was used in the last step .
The same method can deal with the case $\alpha < 0$, where one is in the domain
 of attraction of the {\em trivial} IR fixed point.One gets:
\begin{equation} R(\alpha) = -\int_{-\infty}^0 dz\ exp\left(-\frac{z}{\alpha}
\right) \frac
{exp\left(-\frac{\beta_1}{\beta_0} z\right)}
{\left(1 - \frac{z}{z_n}\right)^{1+\delta}} \end{equation} 
i.e. the analytic continuation of $R_{PT}(\alpha)$ to $\alpha < 0$ and no power
correction.Note that  Eq.~(4) is not the analytic continuation of  Eq.~(7) ,
which suggests that $R(\alpha)$ is given by two different analytic expressions
according whether $\alpha$ is in the  domain of attraction of the trivial or of 
the non-trivial IR fixed point.

That this feature is quite general has been shown in~\cite{du} , where it was 
observed
that $R(\alpha)$ in  Eq.~(1) satisfies the inhomogeneous differential equation:
\begin{equation} R(\alpha)+\frac{1}{n}{\beta(\alpha)}\frac{dR}{d\alpha}=
\alpha\end{equation}
where $\beta(\alpha)=\frac{d\alpha}
{d ln Q^2}$ is a general beta function.The solution is :
\begin{equation} R(\alpha) = C(\alpha) \exp\left(-n\int^{\alpha}\frac{dx}
{\beta (x)}\right)
\end{equation}
where the second factor is the solution of the homogeneous equation , and:
\begin{equation} C(\alpha) = \int_{\alpha_0}^{\alpha}n \frac{dx}{\beta(x)}x
\exp\left(n\int^x
\frac{dy}{\beta(y)}\right)\end{equation}
This result can also been obtained by performing the change of variable 
$k \rightarrow \alpha_{eff}(k)$ in the defining integral Eq.~(1) , which shows 
that $\alpha_0 = \alpha_{eff}(k=0)$ . Consequently~, $R(\alpha)$ does
indeed take 
two different analytic forms , according whether $\alpha_0 = 0$ or $\alpha_0 = 
\alpha_{IR} $ , corresponding to the initial value $\alpha$ being on one side
 or another of the ``separatrix''~\cite{du} in the complex $\alpha$ plane
(more generally, one can consider the case of an arbitrary value  
$0<\alpha_0 < \alpha_{IR}$ , which corresponds to put an IR cut-off in the 
integral Eq.~(1) at $k=k_{min}>0$ with $\alpha_{eff} (k=k_{min})=\alpha_0$) . 
Since two solutions of Eq.~(8) differ by a solution of the homogeneous equation,
which is just a power correction, one gets:
\begin{equation} R(\alpha) = R_{PT}(\alpha) + C \left(\frac{\Lambda^2}{Q^2}
\right)^n \end{equation}
where $R_{PT}(\alpha)$ is the solution corresponding to $\alpha_0 = 0$ ,
which , 
as we have seen in a peculiar case above (Eq.~(7)) , can be shown to be given by
the Borel sum :
\begin{equation} R_{PT}(\alpha) = \int_0^{\infty} dz\ exp\left(- \frac{z}
{\alpha}\right) 
R(z) \end{equation}

To compute the normalization $C$ of the power correction in Eq.~(11) , it is 
useful to transform Eq.~(8) to Borel space~\cite{g}:
\begin{equation} R(z) - \frac{\beta_0}{n}\ z\ R(z) - \frac{1}{n} \int_0^z dy\
 b(z - y)
\ y\ R(y) = 1 \end{equation}
where $b(z)$ is the Borel image of  $b(\alpha) \equiv - \frac{\beta_{eff}
(\alpha)}{\alpha^2} - \beta_0$. $R(z)$ is the solution of Eq.~(13) with
the boundary conditions : $R(z=0) = 1$ and $\frac {dR} {dz}(z=0) = \frac {1} 
{z_n}$.
The crucial observation (made independently in~\cite{du}) is that $R(\alpha)$ 
remains {\em finite} (and approaches $\alpha_{IR}$) for $Q^2 \rightarrow 0$
, as 
is clear from Eq.~(1) .  Eq.~(11) then implies that $R_{PT}(\alpha) \sim
- C \left(\frac{\Lambda^2}{Q^2}\right)^n \rightarrow \infty $ for $Q^2 
\rightarrow 0$ . This behavior can be reproduced with the following ansatz for 
the ``strong coupling" ($z \rightarrow +\infty $) behavior of $R(z)$ 
(assuming $\alpha_{IR}$ is a simple zero of $\beta (\alpha)$) . Put :
\begin{equation} R(z) \equiv \exp\left(\frac {z} {\alpha_{IR}}\right) F(z) 
\end{equation} 
and assume :
\begin{eqnarray} F(z) & \sim & \frac {K} {z^a}\nonumber\\
                 z   & \rightarrow & +\infty \end{eqnarray}
with $\em a<1$ . It is easy to show from the Borel representation Eq.~(12) 
these conditions indeed imply : 
 
\begin{eqnarray} R_{PT}(\alpha) & \sim & K\Gamma (1-a) \left (\frac {\Lambda^2}
{Q^2} \right)^{\omega(1-a)}\nonumber\\
Q^2 & \rightarrow & 0 \end{eqnarray}
where $\omega$ is the "critical exponent`` :
\begin{eqnarray} \beta(\alpha) & \sim & \omega (\alpha - \alpha_{IR})\nonumber\\
 \alpha & \rightarrow & \alpha_{IR} \end{eqnarray}
 Eq.~(16) determines $C = -K\Gamma (1-a)$  from the large $z$ behavior of
$R(z)$ 
and reveals that $n = \omega(1-a)$ .

The connection of the power correction with renormalons stems from the 
observation that K is 
in general {\em complex} : $K = (-1)^{\delta}K_{sing} + K_{reg}$  where the two
real constants $K_{sing}$ and $K_{reg}$ normalize the asymptotic behavior of the
singular "renormalon part`` $R_{sing}(z)$ and of the regular part $R_{reg}(z)$
respectively , defined by :
\begin{eqnarray} R(z) & = & \exp\left(\frac {z} {\alpha_{IR}}\right) F_{sing}(z)
 +
\exp\left(\frac {z} {\alpha_{IR}}\right) F_{reg}(z)\nonumber\\
                      & \equiv & R_{sing}(z) + R_{reg}(z) \end{eqnarray}

where 
\begin{eqnarray} F_{sing}(z) = \left (\frac {1} {1-\frac {z} {z_n}}\right)
^{1+\delta}\times (entire\  function) & \sim & (-1)^{\delta}\frac {K_{sing}} 
{z^a}\nonumber\\  
   z & \rightarrow & +\infty \end{eqnarray}
and
\begin{eqnarray}  F_{reg}(z) = (entire\  function) & \sim & 
\frac {K_{reg}} 
{z^a}\nonumber\\  
   z & \rightarrow & +\infty \end{eqnarray}
The asymptotic behavior for $Q^2 \rightarrow 0$ of the corresponding
 Borel integrals $R_{PT,sing}(\alpha)$ and $R_{PT,reg}(\alpha)$ is then given by
 the analogue of Eq.~(16) , with $K$ replaced by $(-1)^{\delta}K_{sing}$ and
 $K_{reg}$ respectively.
Consequently , one expects the cancellation of the renormalon ambiguity to be 
implemented through the identity (for simplicity I assumed $\delta <0$ for 
convergence of the integral at $z=z_n$) :

\begin{equation} \int_{z_n}^{\infty} dz\ exp\left(-\frac{z}{\alpha} \right) 
R{sing}(z) 
\equiv (-1)^{\delta}K_{sing}\Gamma (1-a) \left (\frac {\Lambda^2}
{Q^2} \right)^n \end{equation}
One deduces the following  general reprentation or $R(\alpha)$ :
\begin{eqnarray} R(\alpha) & = & \int_0^{\infty} dz\ exp\left(- \frac{z}
{\alpha}\right) 
R(z) - \left ((-1)^{\delta}K_{sing} + K_{reg} \right )\Gamma (1-a) \left
 (\frac {\Lambda^2}{Q^2} \right)^n \nonumber\\
                           & = & \int_0^{z_n} dz\ exp\left(- \frac{z}{\alpha}
\right) R_{sing}(z) + \nonumber \\
                           &   & \int_0^{\infty} dz\ exp\left(- \frac{z}{\alpha}
\right) 
R_{reg}(z) - K_{reg} \Gamma (1-a) \left (\frac {\Lambda^2}{Q^2} \right)^n 
\end{eqnarray}
where the last two terms are unrelated to renormalons.The "finite support`` 
Borel representation therefore holds in general only if $R_{reg}(z)\equiv 0$ 
(which 
happens for the two-loop $\beta$ function) . As noted in~\cite{du} , it is also 
possible that $R_{sing}(z)\equiv 0$ (e.g. if $\delta$ is a negative integer ) ,
in which case only the last two terms in Eq.~(22) are present . These results 
have been checked in the 3-loop $\beta$ function case $\beta (\alpha) = 
 - \beta_0 (\alpha)^2 - \beta_1 (\alpha)^3 - \beta_2 (\alpha)^4 $ where one 
can show that  $F(z)$ in Eq.~(14) , if considered as a function of the variable
$x = \frac {\omega} {\beta_0 \alpha_{IR}^2}(z_n - z)$ , satisfies the confluent
hypergeometric equation :
\begin{equation} x \frac {d^2F} {dx^2} + (c-x) \frac {dF} {dx} - aF = 0 
\end{equation}
with $c=2 + \delta $ .

To conclude , it appears that IR renormalons do not necessarily reflect the 
existence of a Landau singularity for {\em physical}\footnote{ A Landau
 singularity may however be present for {\em negative or complex} 
 initial values of the coupling~: in the two loop example the corresponding 
renormalization group trajectory coincides with the "separatrix``.} 
\ initial values of the coupling,
and therefore are not necessarily a signal for non-perturbative physics . They 
do however imply the inadequacy of the Borel summation procedure , which has to 
be amended by a power correction.The latter appears difficult to
distinguish from
"genuine`` power corrections of truly non-perturbative origin (such as QCD sum
rules "condensates``~\cite{svz})~. One attractive possibility is to 
assume~\cite{svz} that "genuine``  power corrections are much larger , in some 
relevant energy range where perturbative QCD still applies , then the purely 
perturbative radiative corrections : such a "mismatch`` is probably not possible
with the type of perturbatively generated power corrections here investigated.
This feature would also make  a (perturbatively) IR 
finite effective coupling an attractive starting point to clarify what is meant 
 by "perturbative `` \ versus \ "power corrections~`` contributions to a process
 (see  the approach of ~\cite{{dw},{dmw}}) .
\section*{Acknowledgements} I thank Yu.L. Dokshitzer and N.G. Uraltsev for 
enlightning discussions.

This research was supported in part by the EC program `Human Capital and 
Mobility',
Network `Physics at High Energy Colliders' , contract CHRX-CT93-0357 
(DG12 COMA).

\end{document}